\documentclass[runningheads]{llncs}
\usepackage{butterma}
\idline{The final publication is available at
\href{http://www.springerlink.com}{www.springerlink.com}\\
L.~Aroyo et al.\ (Eds.): ESWC 2010, LNCS 6089}
\renewcommand{\year}{2010}

\newif\ifdraft
\draftfalse

\ifdraft
\usepackage[show]{ed/ed}
\usepackage[eso-foot,today]{svninfo}
\svnInfo $Id: gencs-lod.tex 1800 2010-03-10 21:21:22Z clange $
\svnKeyword $HeadURL: https://svn.omdoc.org/repos/jomdoc/doc/pubs/eswc-demo10/gencs-lod.tex $
\else
\usepackage[hide]{ed/ed}
\fi

\usepackage{xspace}
\usepackage{paralist}
\usepackage{amsmath}
\usepackage{amssymb}
\usepackage{graphicx}
\usepackage{ifthen}
\usepackage[svgnames]{xcolor}

\definecolor{NavyBlue}{cmyk}{0.94,0.54,0,0.3}
\usepackage[pdfpagescrop={91 71 521 721},a4paper=false,pdftex,pdfstartview=FitV,plainpages=false,pdfpagelabels,colorlinks=true,linkcolor=NavyBlue,citecolor=NavyBlue,urlcolor=NavyBlue,hypertexnames=true]{hyperref}
\hypersetup{
    pdfauthor = {Catalin David and Michael Kohlhase and Christoph Lange and Florian Rabe
    and Nikita Zhiltsov and Vyacheslav Zholudev},
    pdftitle = {Publishing Math Lecture Notes as Linked Data},
    pdfkeywords = {Semantic Web Linked Data Lecture Notes LaTeX sTeX OMDoc OpenMath MathML RDFa
    Mathematics LaTeXML TNTBase Krextor Mocassin JOMDoc JOBAD}
}

\newcommand{\activemath}{ActiveMath\xspace}

\newcommand{\mathml}{MathML\xspace}

\newcommand{\omdoc}{OMDoc\xspace}
\newcommand{\openmath}{OpenMath\xspace}
\newcommand{\rdfa}{RDFa\xspace}

\newcommand{\xhtml}{XHTML\xspace}

\title{Publishing Math Lecture Notes as Linked Data}
\author{Catalin David\inst{1}, Michael Kohlhase\inst{1}, Christoph Lange\inst{1}, Florian
Rabe\inst{1}, Nikita Zhiltsov\inst{2} and Vyacheslav Zholudev\inst{1}}
\institute{Computer Science, Jacobs University Bremen, Germany
\email{\{c.david,m.kohlhase,ch.lange,f.rabe,v.zholudev\}@jacobs-university.de} \and Mathematics, Kazan State University, Russia, \email{nikita.zhiltsov@gmail.com}}
\authorrunning{David, Kohlhase, Lange, Rabe, Zhiltsov, Zholudev}

\begin{document}

\maketitle
\thispagestyle{electronic}

\begin{abstract}
  We mark up a corpus of {\LaTeX} lecture notes semantically and expose them as Linked Data in {\xhtml}+\linebreak[1]{\mathml}+\linebreak[1]{\rdfa}.   Our application makes the resulting documents interactively browsable for   students.  Our ontology helps to answer queries from students and lecturers, and paves   the path towards an integration of our corpus with external sites.
\end{abstract}

\section{Application: Computer Science Lecture Notes}
\label{sec:application}

Over the last seven years, the second author has accumulated a large corpus of teaching materials, comprising more than 2,000 slides, about 1,000 homework problems, and hundreds of pages of course notes, all written in {\LaTeX}. The material covers a general first-year introduction to computer science, graduate lectures on logics, and research talks on mathematical knowledge management.
This situation is typical for educators and researchers and represents the state of the art in mathematics, physics, computer science, and engineering%
: {\LaTeX} has proven suitable for writing high-quality lecture notes and publishing them as PDF.  However, in our educational setting, we would like to benefit from the much larger degree of interactivity that screen reading and e-books support.  For example, while reading notes students want to directly look up the meaning of a symbol (e.\,g.\ $\vDash$) in a formula, or examples for a difficult concept (e.\,g.\ structural induction).  They may want to select advanced material for self-study from the whole body of lecture notes, based on the topics covered in the lecture.  They want to use a search engine to find related material in other universities' online course notes, on mathematical web sites, or Wikipedia.  Lecturers want to query their repository for document parts  reusable in an upcoming lecture, given the prerequisites students are expected to meet and the material that has already been covered.  In a course for a special audience, e.\,g.\ mathematics for physicists, they want to draw examples from that domain even though they are less familiar with it.  They also want to locate didactic gaps, such as concepts without examples, or unjustified proof steps.  These services require semantic annotations in the lecture notes that are understandable for external search engines. %
Plain {\LaTeX} is barely usable for anything \emph{beyond} on-screen reading and printing.  Even simple semantic annotations are uncommon, rare exceptions are the \verb|\title| command making its meaning explicit or \verb|\frac{a}{b}| focusing on functional structure instead of visual layout.  This is especially problematic for symbols in formul\ae, which are often overloaded with multiple definitions or presentable in different notations.  $\binom{n}{k}$ can be a vector or a binomial coefficient, and a French or Russian would write the latter as $\mathcal{C}^k_n$. %
Therefore, we have developed a semantic representation of mathematical knowledge in {\LaTeX} and a presentation process that preserves these semantic structures as Linked Data in the output, exposing them to mashups for interactive exploration, as well as semantic searching and querying.  These are based on an ontology for mathematical knowledge so that mathematical content can be linked across different repositories.
 
\section{Research Background and Related Work}
\label{sec:research}

{\LaTeX}'s importance in scientific authoring and its extensibility by macros have led to semantic extensions enabling modern publishing workflows.  SALT (semantically annotated {\LaTeX}~\cite{Groza:SALT07}) marks up rhetorical structures and fine-grained citations in scientific documents.  Its markup is not sufficiently fine-grained for formul\ae, and its vocabulary is limited to rhetorics and citations and not extensible.   Our own s{\TeX} offers macros for introducing new mathematical symbols and using arbitrary metadata vocabularies.
  Some math e-learning systems, such as ActiveMath~\cite{activemath:web} or MathDox~\cite{mathdox:web}, use semantic representations of formul\ae\ and higher-level structures, e.\,g.\ proof steps or course module dependencies, in the standard XML languages OpenMath~\cite{BusCapCar:2oms04} and OMDoc~\cite{Kohlhase:omdoc1.2}. They utilize semantic structures  but do not \emph{publish} them in a standard representation like RDF, which would promote general-purpose queries beyond the built-in services and integration with other systems on the web.  The Linking Open Data movement promotes best practices for publishing data on the web~\cite{LinkedDataGuidesTutorials:web}, as standalone RDF or embedded into HTML documents as RDFa~\cite{AdidaEtAl08:RDFa}.  Applications include Sindice, an engine that crawls and indexes Linked Data~\cite{sindice:web}, and the Sparks $O_3$ Browser, a mashup that utilizes RDFa annotations in HTML for interactive browsing~\cite{SparksOzone:web}.  Our interactive documents work similarly but additionally support annotations in MathML formul\ae.  MathML has pioneered embedded annotations long before RDFa, albeit with a more limited scope.  Its \emph{parallel markup} interlinks the rendered appearance and the semantic structure of mathematical expressions; the meaning of mathematical symbols is usually defined in lightweight ontologies called OpenMath content dictionaries~\cite{W3C:MathML3:biblatex}.  HELM (Hypertext Electronic Library of Mathematics~\cite{APSGS:MKM-HELM03}) pioneered representing structures of mathematical knowledge in RDF, e.\,g.\ what mathematical theory introduces a symbol, what of its properties have been declared or asserted, and how the latter are proved.  The HELM ontology has not gained wide acceptance, though. At the time of its development, there was no RDFa-like way of embedding RDF into web documents.

\section{Architecture and Demo}
\label{sec:technology}

Our architecture publishes semantically enriched {\LaTeX} lecture notes as XHTML+\linebreak[1]MathML+\linebreak[1]RDFa Linked Data.  We kept {\LaTeX} as an input language, as it is familiar to authors and well supported by editors, and as high-quality PDF can be obtained from it. With s{\TeX} (semantically enhanced {\TeX}), we have introduced {\LaTeX} macros for marking up the semantic structure of formul\ae\ and documents~\cite{Kohlhase:ulsmf08} and manually annotated our complete corpus using the s{\TeX} plugin for the Emacs editor.  One can, e.\,g., declare a symbol \textit{union}, formally define it, and make its semantic representation \verb|\union{A,B,C}| expand to \verb|A\cup B\cup C| for human-readable rendering.  There are environments for mathematical statements and theories, e.\,g.\ \verb|\begin{example}[for=union]|.     {\LaTeX}ML transforms this into a semantically equivalent intermediate XML representation, using the standard XML languages OpenMath for   formul\ae~\cite{BusCapCar:2oms04} and OMDoc for higher-level   structures~\cite{Kohlhase:omdoc1.2}.  Finally, our JOMDoc rendering   library~\cite{JOMDoc:web} generates human-readable output from this XML -- an output   that still contains the full semantic structure as annotations.  A custom Java   implementation renders formul\ae\ as parallel markup of Presentation MathML annotated   with OpenMath\footnote{A proposal for fully representing formul\ae\ in     RDF~\cite{Marchiori:MathematicalSemanticWeb03} has not gained wide acceptance.     RDF-based reasoners are often limited to decidable first order logic subsets, which is     insufficient for mathematical applications, and XML has a straightforward     notion of order (e.\,g.\ of the arguments of an operator or of a set     constructor).}; rendering higher-level structures as   XHTML+RDFa~\cite{AdidaEtAl08:RDFa} is implemented in XSLT.  RDF is extracted from XML   by our Krextor XML$\to$RDF library~\cite{Lange:Krextor09}, which generates URIs for all mathematical objects in a document.  It uses our OMDoc   ontology (cf.~\cite{lange:swim-demo08}) as a vocabulary for representing all mathematical structures (e.\,g.\ ``$d$ is a definition, $e$ is an example for $d$'') plus full text, inspired by   HELM and designed as a more expressive counterpart of the OMDoc XML schema.

The whole transformation process is integrated into our versioned XML database TNTBase~\cite{ZhoKohRab:tntbasef10}; see \url{http://kwarc.info/LinkedLectures}.  TNT\-Base has a Subversion-compatible interface making it suitable as a lecture notes repository.  The {\TeX}$\to$XML and XML$\to$RDF transformations are automatically triggered by a hook upon committing a new revision of an s{\TeX} lecture module.  If the generated OMDoc+OpenMath is not schema-valid, the commit is rejected.  On the other hand, it follows Linked Data best practices and, depending on the MIME type an HTTP client requests, serves a document as OMDoc, as RDF (only a structural outline, not the full text and formul\ae), or as XHTML+\linebreak[1]MathML+\linebreak[1]RDFa. The latter contains JavaScript code from our JOBAD library for interactive documents~\cite{KGLZ:JOBADabstract09,GLR:WebSvcActMathDoc09}, which operationalizes the annotations -- Linked Data and other -- in the rendered documents.  JOBAD's definition lookup determines the OpenMath annotation of the Presentation MathML symbol the user clicked on, from that obtains the URI of the symbol%
, and then requests XHTML from that URI (resulting in the symbol's declaration and definition), which is then displayed in a popup.  The RDFa annotations are used for making parts of a document (e.\,g.\ steps of a structured proof) foldable, and for displaying the local neighborhood in the RDF graph (e.\,g.\ related examples) in popups; this is implemented using the rdfQuery library~\cite{rdfQuery:web}, relying on the Linked Data structure in the latter case.  Further third-party services can be integrated in a mashup style; we have demonstrated this for a unit conversion service~\cite{KGLZ:JOBADabstract09,GLR:WebSvcActMathDoc09}.  Besides  enabling JOBAD's services, we 
have implemented machinery to load the extracted RDF into a triple store and query it using SPARQL. 
We also provide a widget for formulating queries without knowing SPARQL and the OMDoc ontology. It allows to ask some non-trivial queries, e.\,g.\ ``find examples for all concepts from graph theory (about which I'm planning a lecture), assuming as prerequisites the concepts from formal languages (and their prerequisites)''.  This would yield the parse tree of a context-free language as an example for the concept ``tree'' -- as operating systems were not among the prerequisites.

\noindent\includegraphics[width=\textwidth]{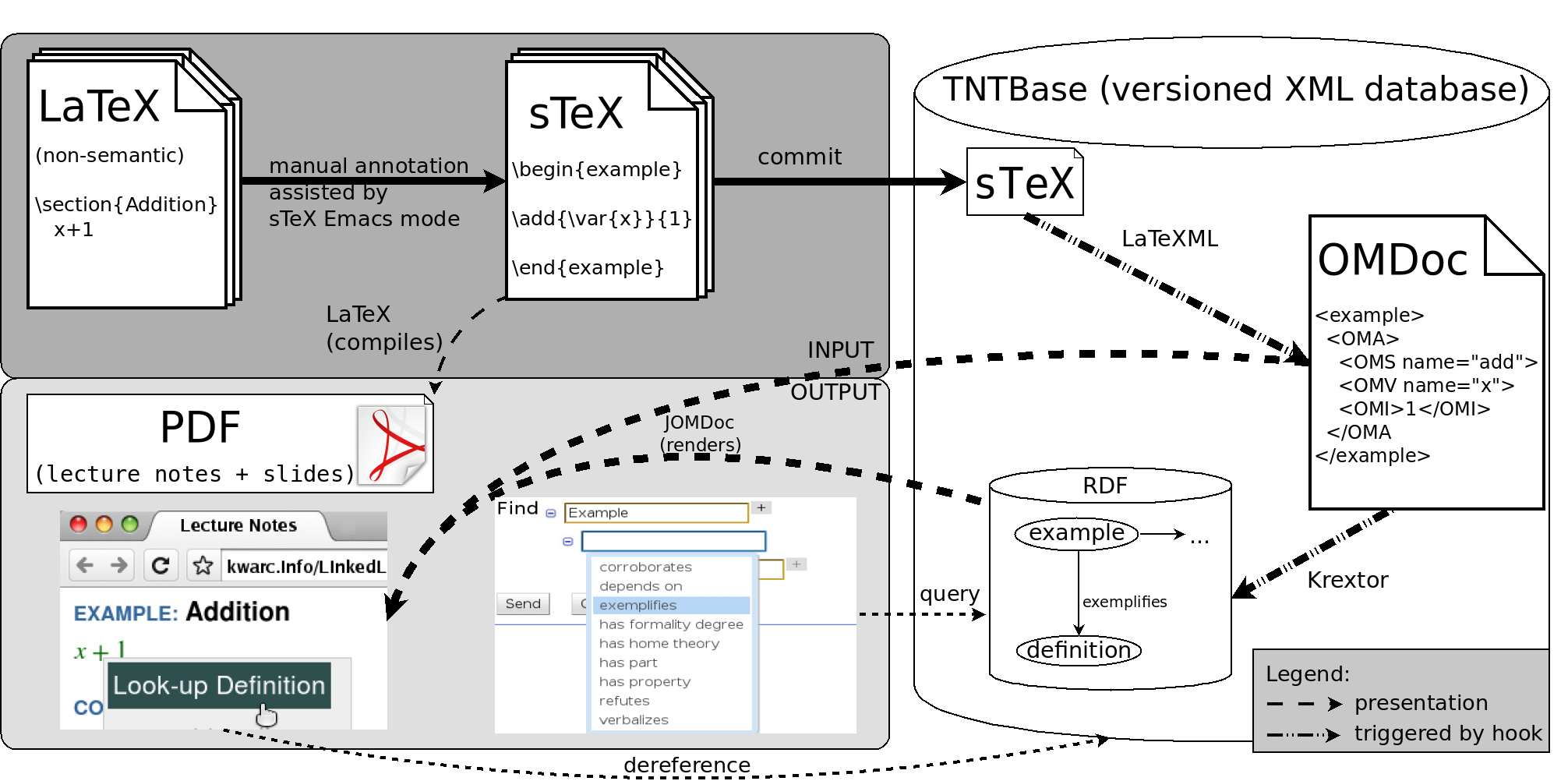}

\noindent Our demo shows the complete pipeline in action: \begin{inparaenum}[(i)]\item annotating   a document with our s{\TeX} Emacs mode, \item committing it to TNTBase, \item   automatic translation to OMDoc, schema validation, and RDF extraction,\item loading the extracted RDF data into a triple store, \item   retrieving the document in different representations, \item browsing the   XHTML+MathML+RDFa rendering, \item interacting with the Linked Data in   it, \item and querying a triple store\end{inparaenum}.  Additionally, we will demonstrate the generation of PDF from the s{\TeX} sources.

\section{Conclusion and Outlook}
\label{sec:conclusion}

Our architecture makes legacy {\LaTeX} lecture notes available as Linked Data.  We expose these data to external clients but have also implemented services for interactively exploring the XHTML+\linebreak[1]MathML+\linebreak[1]RDFa presentation of our data.  We are also working on preserving some of the semantics in the PDF output, as SALT does.  Evaluation of our enriched lecture notes by the student end users is planned for the next semester.  To the best of our knowledge, we are the first provider of RDF-based Linked Data in the domain of mathematics and among the first to operationalize the Linked Data structures of formula markup.  Having successfully transformed more than 300,000 normal, non-semantic {\LaTeX} documents from \url{arxiv.org} to XHTML+Presentation MathML~\cite{StaKoh:tlcspx09} and working on machinery for automatically annotating them using natural language processing, we will soon be able to expose even more mathematical knowledge as Linked Open Data; however, due to the inherent complexity of mathematical knowledge, with a less formal semantics than manually annotated documents.  Our lecture notes are self-contained so far, but we are now starting to reap the benefits of Linked Data by linking them to  other data sets, e.\,g.\ DBpedia~\cite{dbpedia:web}, whose mathematical knowledge does not have a semantics as strong as ours, but which provides abundant informal background knowledge, e.\,g.\ about the originators of mathematical theories.  On the other hand, hardly any well-known mathematical site (e.\,g.\ \url{planetmath.org} and \url{mathworld.wolfram.com}) currently exposes machine-understandable metadata.  We promote our technology, starting with lightweight RDFa annotation using the OMDoc ontology, as a migration path towards their integration into a true mathematical Semantic Web.

\providecommand\seen{seen } \providecommand\webpageat{web page at }
  \providecommand\homepageat{home page at }
  \providecommand\projectpageat{project page at }
  \providecommand\systempageat{system home page at }
  \providecommand\svnrepoat{Subversion repository at }
  \providecommand\January{January} \providecommand\February{February}
  \providecommand\March{March} \providecommand\April{April}
  \providecommand\May{May} \providecommand\June{June}
  \providecommand\July{July} \providecommand\August{August}
  \providecommand\September{September} \providecommand\October{October}
  \providecommand\November{November} \providecommand\December{December}
  \providecommand\AUSTRALIA{Australia} \providecommand\ROMANIA{Romania}
  \providecommand\MEXICO{Mexico} \providecommand\ITALY{Italy}
  \providecommand\USA{USA} \providecommand\IRELAND{Ireland}
  \providecommand\HUNGARY{Hungary} \providecommand\JAPAN{Japan}
  \providecommand\CANADA{Canada} \providecommand\SPAIN{Spain}
  \providecommand\NETHERLANDS{Netherlands} \providecommand\UK{UK}
  \providecommand\SWEDEN{Sweden} \providecommand\GERMANY{Germany}
  \providecommand\openmath{OpenMath} \providecommand\fc{forthcoming}
  \providecommand\PROC{Proceedings} \providecommand\omdoc{OMDoc}
  \providecommand\activemath{ActiveMath}

\end{document}